\documentstyle[epsfig,12pt]{article}
\def\ka14{$K^*_0(1430)$\ }
\def\d3pi{$D^+\rightarrow \pi^-\pi^+\pi^+$\ }
\def\ds3pi{$D_s^+\rightarrow \pi^-\pi^+\pi^+$\ }
\def\f2{$f_2(1270)$\ }

\begin{document}

\title{    Phase Motion in the Scalar Low-Mass $\pi \pi$ Amplitude in
                $D^+ \to \pi^- \pi^+ \pi^+$ Decay }

\author{Ignacio Bediaga and Jussara M. de Miranda \\ Centro
Brasileiro de Pesquisas F\'\i sicas, \\ Rua Xavier Sigaud 150, 22290-180
-- Rio de Janeiro, RJ, Brazil\\ bediaga@cbpf.br and jussara@cbpf.br}
\maketitle

\begin{abstract}

Applying  the Amplitude Difference  method to Fermilab experiment E791   
$D^+ \to \pi^-\pi^+\pi^+$  data,
we measure the
low mass $\pi^+\pi^-$   phase  motion.  Our results suggest a significant 
phase variation, compatible with the existence of an isoscalar  
$\sigma(500)$ meson, as previously reported  using 
an isobar model fit to the full Dalitz-plot density.

\end{abstract}

\vspace{3cm} 
\begin{tabbing}

\=xxxxxxxxxxxxxxxxxx\= \kill

\>{\bf Submitted to Physics Letters B} \>  \\

\>{\bf Keywords:} Heavy Meson Decays; Dalitz-plot; Scalar Mesons; Resonances.\>  \\

\>{\bf PACS Numbers:} \> 13.25.Ft, 14.40.Ev, 14.40.Cs,11.80.Et.

\end{tabbing}

\newpage

 Recently we proposed the isobar-based Amplitude Difference (AD) method  to extract  
the phase motion of a complex amplitude in three-body heavy-meson 
decays \cite{ad}. With this method, the 
phase variation of a generic complex amplitude can be directly revealed through 
interference  in the Dalitz-plot region where it crosses  a well
established  resonant state, used as a probe. 
 As a test, this method was successfully applied to data \cite{UT} to  extract  the  
well known  phase motion  of   the scalar amplitude $f_0(980)$  observed 
in  \ds3pi\footnote{ Charge conjugate states are implied throughout the paper.} decay. In the 
present paper we  use the same method to study the
low $\pi^+\pi^-$ mass region of the \d3pi decay where  Fermilab experiment 
E791 showed evidence for the existence of a light and broad 
scalar resonance \cite{prl}.

 To obtain  good  
fit quality in a full Dalitz-plot analysis, E791  found it necessary to  include 
an  extra scalar particle, in 
addition to the well-established di-pion  resonances \cite{pdg}. For this 
new scalar state, parameterized as an S-wave Breit-Wigner resonance, 
they measured a mass and a width of  $478^{+24}_{-23} \pm 17 $ MeV/$c^2$ and  
$324^{+42}_{-40} \pm 21 $ MeV/$c^2$, respectively. These parameters are 
compatible with those expected for the isoscalar meson $\sigma(500)$. 
The $D^+\to\sigma(500)\pi^+$\cite{prl} decay   appeared as the  dominant contribution,
 accounting for  approximately half the \d3pi  decays.

The E791 result has been widely discussed \cite{discussion}-\cite{polosa} 
and new data  has become available \cite{ex_evid}.  However it is 
desirable to be able to confirm the result through a direct
observation of the phase motion expected for a 
resonance \cite{torn1,barnes,ochs}.  In this context  we apply the 
AD method to the low $\pi^+\pi^-$ mass region of the  \d3pi decay. 
We also compare the phase variation of the 
Breit-Wigner function found in the isobar Dalitz-plot analysis \cite{prl} to the 
model-independent method of this paper.

\begin{figure}[t]
\epsfxsize=18pc
\centerline{
\epsfbox{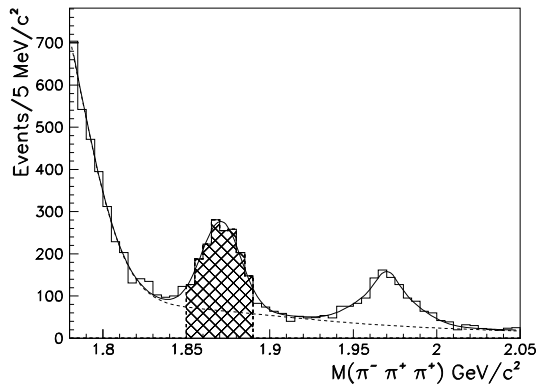}}
\caption{The $\pi^-\pi^+\pi^+$ invariant mass spectrum. The  dashed line
represents the total background. Events used for the $D^+$ isobar Dalitz-plot 
analysis [3] and in this AD method analysis  are in the hatched area.}
\label{m3pi}
\end{figure}

The present study is a reanalysis of  the  Fermilab experiment E791 data. 
 Here we investigate a subset of the total phase space  used by the experiment in their full  
 Dalitz-plot analysis. A description of the experiment, data selection criteria,
 background parametrization and detector acceptance are found in references
  \cite{prl,prlds}. 
The final $\pi^-\pi^+\pi^+$ invariant mass distribution is shown in
Fig. \ref{m3pi}. There are  1686  events with invariant mass between 
1.85 and 1.89 GeV$/c^2$ shown  in the shaded region of Fig. \ref{m3pi}. 
 The integrated signal-to-background ratio in this range is about 2:1. Fig.
2 shows the folded Dalitz-plot.
The horizontal  and vertical axes are the squares of the $ \pi^+ \pi^- $ 
 invariant mass high ($s_{12}$) and low ($s_{13}$) combinations.
 The  analysis presented here uses  the hatched area of Fig. 2. 
 We estimate 60  background events  in a total of  197 candidate events.
 The background does not show any dependence on the $s_{12}$ variable \cite{prl}.
% Based in the background
% study done for the previous analysis, we find that its distribution shows no
% dependence in the $s_{12}$ variable.
  
  The detector acceptance in this region is almost constant. There is a very mild
  slope in the $s_{13}$ acceptance, but no variation with $s_{12}$. Nevertheless the
  acceptance is  taken into account to correct the event
  distributions shown later.

%%%%%%%%%%%%%%%%%%%%%%%%%%%%%%%%%%%%%%%%%%%%%%%%%%%%%%%%%%%%%%%%%%%
%%%%%%%%%%%%%%%%%%%%%%%%%%%%%%%%%%%%%%%%%%%%%%%%%%%%%%%%%%%%%%%%%%%

%%%%%%%%%%%%%%%%%%%%%%%%%%%%%%%%%%%%%%%%%%%%%%%%%%%%%%%%%%%%%%%%%%%
\begin{figure}[hbt]
\centerline{\epsfysize=3.0in \epsffile{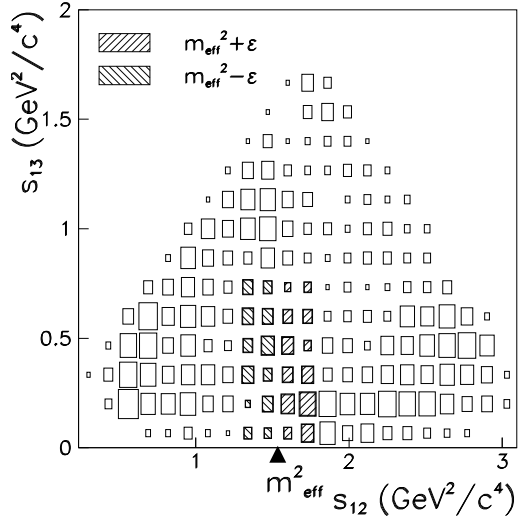}}
\caption  { The folded $ D^+\to \pi^-\pi^+\pi^+ $ Dalitz-plot distribution of the events
in the hatched area of Figure 1.  
The events used by the AD method analysis are in the hatched region of this
folded Dalitz-plot. The size of the
area of each bin in the plot corresponds to the number of events in that bin.}
\label{dalitz}
\end{figure}

%%%%%%%%%%%%%%%%%%%%%%%%%%%%%%%%%%%%%%%%%%%%%%%%%%%%%%%%%%%%%%%%%%%

There are two conditions necessary   to extract
the phase motion  of a generic  amplitude with the AD method:

\begin{itemize}

\item A crossing region between  the amplitude under 
study and a probe resonance  has to be dominated by these two contributions.

\item The integrated amplitude of the probe resonance must be symmetric 
with respect to an  effective mass squared ($m^2_{eff}$).
\footnote{ An analysis with more amplitudes in 
a limited phase space could be done, but in a model dependent way.}  

\end{itemize}

\begin{figure}[hbt]
\centerline{\epsfig{file=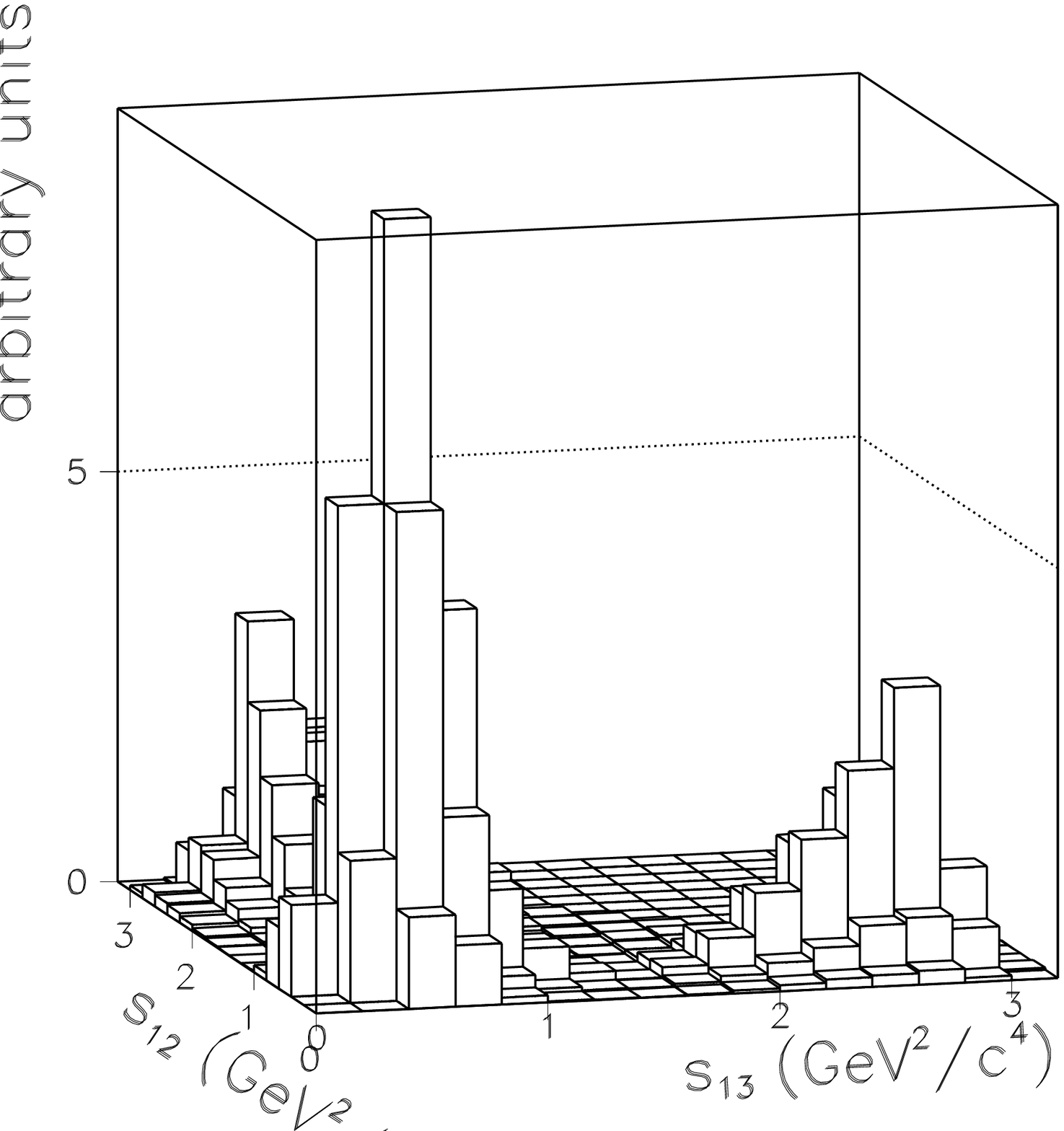,width=6 cm}}
\caption{ Fast MC $\rho(770) \pi$ Dalitz-plot distribution in \d3pi decay.}
\label{rho}
\end{figure}

To study the low mass region in $s_{13}$, 
three  well known resonances could serve as a probe in $s_{12}$ 
 in the $ D^+\to \pi^-\pi^+\pi^+ $ decay: $\rho(770)$, $f_0(980)$ and 
$f_2(1270)$. However the broad $\rho(770)$ and  $f_0(980)$ are too close 
to each other to pass the isolation criteria mentioned above, as can be seen in
 Fig. \ref{dalitz}. 
On the other hand, the tensor $f_2(1270)$, $m_0^2 = 1.61$ GeV$^2/c^4$,
 is located where the $\rho(770)$  reaches a minimum due to its decay 
 angular distribution in the 
crossed ($s_{13}$) channel, (see Fig.~\ref{rho}).  
 In the \d3pi decay, 
the  $f_2(1270)$ contribution satisfies the necessary  conditions of 
 having a substantial  contribution crossing the low   mass region, this being
 the regions where all other amplitudes can be  considered negligible. In particular,
we estimate a contamination of 5\% of $\rho(770)\pi^+$ events in the region 
of interest.

 %%%%%%%%%%%%%%%%%%%%%%%%%%%%%%%%%%%%%%%%%%%%%%%%%%%%%%%%%%%%%%%%%%%
%%%%%%%%%%%%%%%%%%%%%%%%%%%%%%%%%%%%%%%%%%%%%%%%%%%%%%%%%%%%%%%%%%%

 We assume that the only contributions in this region  are the  
$f_2(1270)$ amplitude in  $s_{12}$ and the
  $\pi \pi $  complex amplitude  under study in $s_{13}$. We  write:

\begin{eqnarray}
{\cal A}(s_{12},s_{13}) \approx a_{R}\hspace{.3cm}  {\cal
BW}_{f_2(1270)}(s_{12}) {
\hspace{.2cm}   ^{J=2}{\cal M}_{f_2(1270)}(s_{12},s_{13})} + \\
+\hspace{.3cm} a_s/(p^*/\sqrt s_{13}) \hspace{.2cm} sin \delta (s_{13}) \hspace{.3cm}
e^{i(\delta(s_{13})+\gamma)},
\nonumber
\end{eqnarray}

\noindent where $\gamma$  is the overall relative phase-difference between the two 
amplitudes  (assumed to be constant and arising from both production 
and final-state-interaction (FSI) between the di-pion system and the bachelor
pion); 
$sin\delta (s_{13}) e^{i\delta(s_{13})}$ represents the most general 
amplitude for a two-body elastic scattering, $p^*/\sqrt s_{13}$ is a 
phase space integration factor to make this description compatible with $\pi \pi$ 
scattering and $p^*$ is the  pion momentum measured  in the resonance rest frame;  
$^{J=2}{\cal M}_{f_2(1270)}(s_{12},s_{13})$ is the angular
function for the $f_2(1270)$ tensor resonance  given by 
${ \frac{4}{3}(\mid {\bf p_3} \mid \mid {\bf p_2} \mid)^2 
(3cos^2\theta-1)}$, $\theta$ being the angle between the pions 2 and 3, and $J$
the angular momentum of the resonance; and $a_R$ and $a_s$ are
 the production strengths.
The AD method assumes that the $\pi^+\pi^-$ production is 
constant over the effective mass range considered and $a_R$ and $a_s$ 
are energy independent.  Finally the Breit-Wigner distribution 
is given by:
\begin{eqnarray}
  {\cal BW} = {m_0 \Gamma_0 \over {m_0^2 - s - im_0\Gamma(s)}}. 
\end{eqnarray}

 The   width  is given by $\Gamma(s) = \Gamma_0 \frac{m_0}{m}\left
(\frac{p^*}{p^*_0}\right)^{2J+1} \frac{^{J}F^2(p^*)}{^{J}F^2(p^*_0)}$,
where the central barrier factor is $^{J=2}F = 1/{\sqrt{9+3(rp^*)^2+(rp^*)^4}}$.
 The parameter $r$ is the radius of the resonance ($\sim 3 fm$)\cite{argus}  and 
$p^*=p^*(m)$ is the momentum of  decay particles at mass $m$, measured in 
the resonance rest frame, $p^*_0=p^*(m_0)$, where $m_0$ is the 
resonance mass.

\begin{figure}[hbt]
\centerline{\epsfig{file=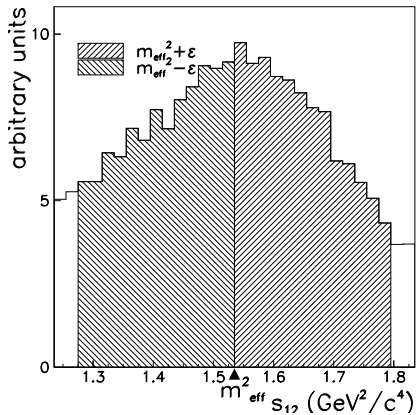,width=6 cm}}
\caption{ Fast MC simulation$f_2(1270)$ $s_{12}$  distribution, divided into
 $m_0^2+\epsilon$ and $m_0^2-\epsilon$, integrated between the 
 threshold and 0.8 GeV$^2$ in  $s_{13}$. }
\label{f2_s12}
\end{figure}

Both $\Gamma(s)$ and  the angular function
  $^{J=2}{\cal M}_{f_2(1270)}$ for the $f_2(1270)$ resonance  produce 
asymmetries in the  $s_{12}$ distribution. We use 
 a  $D^+\to f_2(1270)\pi^+ $ fast Monte Carlo (MC) simulation\footnote{ 
 The Monte Carlo events are  generated
 according to the physical amplitude squared using a uniform phase space density 
 and weighted by the detector acceptance distribution over the Dalitz-plot. } to study the behavior of the 
 probe resonance distribution in the $s_{12}$  and $s_{13}$ variables.  
 The event distribution for $f_2(1270)$ in the $s_{12}$ subsystem from  
 Monte Carlo is shown in  Fig. \ref{f2_s12}. We choose an 
effective square  mass  of $ 1.535$ GeV$^2/c^4$ such that the number of events 
between $ m_{eff}^2$ and $ m_{eff}^2 + \epsilon$ 
($\epsilon = 0.26$ GeV$^2/c^4$) 
 is equal to the number of events integrated between  
 $ m_{eff}^2$ and $ m_{eff}^2 - \epsilon$. 
%%%%%%%%%%%%%%%%%%%%%%%%%%%%%%%%%%%%%%%%%%%%%%%%%%%%%%%%%%%%%%%%%%%
%%%%%%%%%%%%%%%%%%%%%%%%%%%%%%%%%%%%%%%%%%%%%%%%%%%%%%%%%%%%%%%%%%%

\begin{figure}[hbt]
\centerline{\epsfig{file=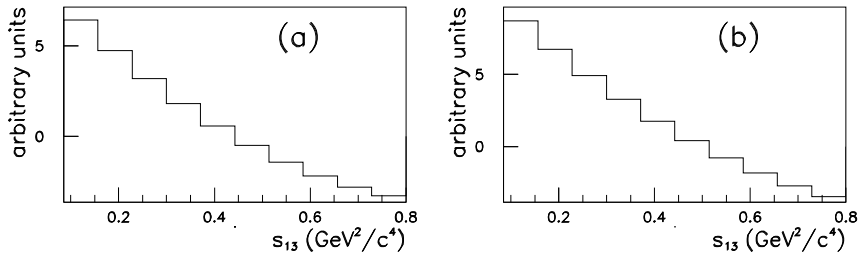,width=13 cm}}
\caption{ Fast MC of angular distribution function alone, $^{J=2}{\cal M}_{f_2(1270)}$,  in $s_{13}$ 
for (a)  events
between  $ m_{eff}^2$ and $ m_{eff}^2 + \epsilon$ and (b)  events
between $ m_{eff}^2$ and $ m_{eff}^2 - \epsilon$.}
\label{f2_s13}
\end{figure}

 For events in $s_{12}$  between $ m_{eff}^2$ and $ m_{eff}^2 + \epsilon$ in
  $s_{12}$, $^{J=2}{\cal M}_{f_2(1270)}(s_{13})$ is shown in  
   Fig. \ref{f2_s13}a and for those between $ m_{eff}^2$ and $ m_{eff}^2 - \epsilon$ 
   in     Fig. \ref{f2_s13}b\footnote{ The plots were produced by a fast MC 
   of the angular distribution function alone, with no detector influence 
   included.   We represent this function with the same  binning  
   used for our data events.}.  
   We can see that  
  these two plots are just slightly different. In our analysis we consider  
  the approximation
$ ^{J=2}{\cal M}^+_{f_2(1270)}(s_{13})\approx  \hspace{.1cm} 
^{J=2}{\cal M}^-_{f_2(1270)}
(s_{13})$  and  take the   average function   
$ ^{J=2}\bar{\cal M}_{f_2(1270)}(s_{13})$.
An important effect that we have to take into account is the zero 
of this function
at $s_{13} \sim $0.48 GeV$^2/c^4$. Below we discuss the consequences 
of that for this AD method application.

 %%%%%%%%%%%%%%%%%%%%%%%%%%%%%%%%%%%%%%%%%%%%%%%%%%%%%%%%%%%%%%%%%%%
%%%%%%%%%%%%%%%%%%%%%%%%%%%%%%%%%%%%%%%%%%%%%%%%%%%%%%%%%%%%%%%%%%%
To be brief, from here on we use
$^{J=2}\bar{\cal M}_{f_2(1270)}(s_{13})= \bar{\cal M} $ and 
$p^*/\sqrt s_{13} = p'$.
The main equation of the AD method applied to the integrated amplitude-square 
difference is 
\cite{ad,UT}:
\begin{eqnarray}
 p' /\bar{\cal M} ( \Delta  \int  {\cal A}^2 )= p' /\bar{\cal M}
(\int_{m_{eff}^2}^{m_{eff}^2 + \epsilon}\mid {\cal A}( s_{12}, s_{13} ) \mid^2 ds_{12}
-\int_{m_{eff}^2 - \epsilon}^{m_{eff}^2}\mid {\cal A}( s_{12}, s_{13}) \mid^2
ds_{12})\nonumber\\
\approx { - {\cal C}(  sin(2 \delta(s_{13})+ \gamma) - sin \gamma),\hspace{2cm}}
\end{eqnarray}

\noindent where ${\cal C}$ is an overall  constant. 

   From  Eq. 2, it follows that the  function $ \Delta  \int { \cal A}^2 p' /\bar{\cal M}$ 
   directly reflects the behavior of $\delta(s_{13})$.  
A constant
 $ \Delta  \int { \cal A}^2 p' /\bar{\cal M}$ implies a constant 
 $\delta(s_{13})$ which is  the case for a   non-resonant 
 contribution. 
 In the same way, a slow phase motion will produce a slowly  varying 
 $ \Delta\int { \cal A}^2 p' /\bar{\cal M}$
   distribution,
  and a full resonance phase motion produces a clear signature in 
  $ \Delta  \int { \cal A}^2 p' /\bar{\cal M}$ with the presence of zero, 
  maximum and minimum values. 
  
\begin{figure}[hbt]
\centerline{\epsfig{file=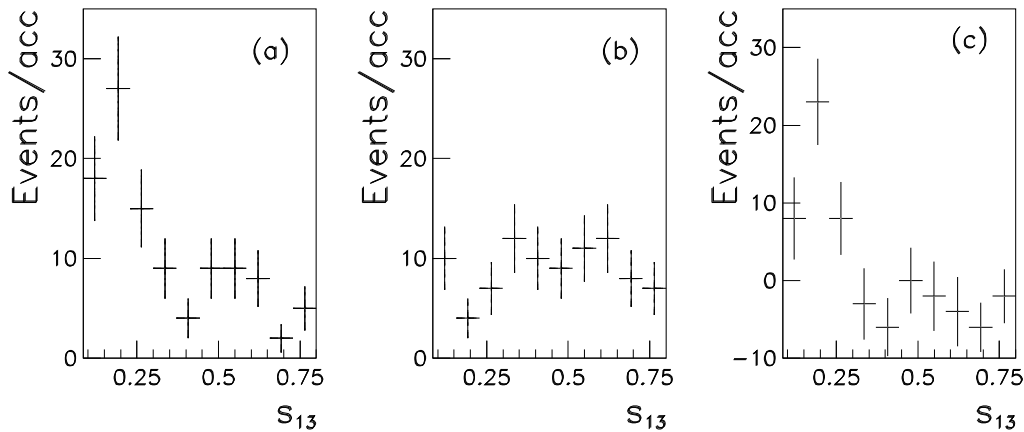,width=15.5 cm}}
\caption{Event  distributions projected onto the  $s_{13}$ axis (a) for all events 
in the $s_{12}$ interval $\int_{m^2_0}^{m^2_0+\epsilon}\mid {\cal A}(s_{12},s_{13})\mid^2
ds_{12}$, and (b) for events ${\int_{m^2_0-\epsilon}^{m^2_0}\mid 
{\cal A}(s_{12},s_{13})\mid^2 ds_{12}}$. The  distributions are acceptance corrected
 such that the overall data statistics is conserved. 
 Plot 
(c) shows the $ \Delta \int  { \cal A}^2 $ distribution .}
\label{data}
\end{figure}

As mentioned previously,  the  background has no dependence on $s_{12}$ and 
its contribution vanishes from   the $ \Delta  \int { \cal A}^2$ distribution. 
The  $ \int { \cal A}^2 $  in $s_{13}$
for events  integrated in $s_{12}$   $m_{eff}^2 $ and
$m_{eff}^2  +  \epsilon $ and $m_{eff}^2$  and $ m_{eff}^2  -  \epsilon $, 
corrected by the acceptance shape, are presented  in  Figs.
\ref{data}a and \ref{data}b, respectively; these events correspond to the 
hatched area of Fig. \ref{dalitz}. We obtain $ \Delta \int { \cal A}^2 $ shown
in histogram \ref{data}c, by 
subtracting  the \ref{data}b histogram from that in \ref{data}a.

 %%%%%%%%%%%%%%%%%%%%%%%%%%%%%%%%%%%%%%%%%%%%%%%%%%%%%%%%%%%%%%%%%%%
%%%%%%%%%%%%%%%%%%%%%%%%%%%%%%%%%%%%%%%%%%%%%%%%%%%%%%%%%%%%%%%%%%%

\begin{figure}[hbt]
\centerline{\epsfig{file=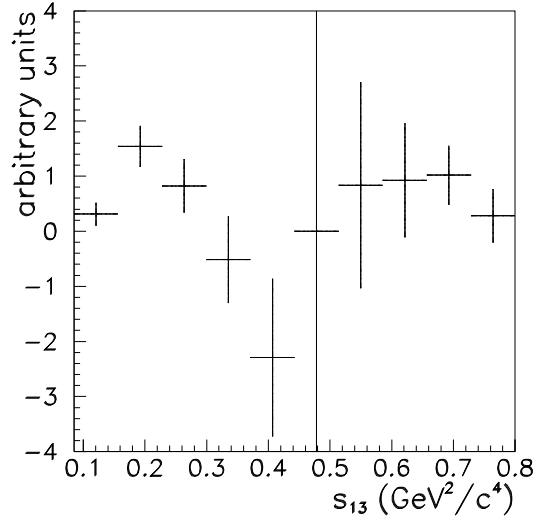,width=8 cm}}
\caption{ The distribution  of 
$ \Delta \int  { \cal A}^2 \hspace{.1cm} p'/ \bar{\cal M} $ in $s_{13}$. Note that 
the scale is arbitrary once 
the data, $ \Delta \int  { \cal A}^2$, is multiplied by an unnormalized 
function $p'/ \bar{\cal M} $.  The error bars represent statistical errors only.}
\label{data3}
\end{figure}

To extract the phase motion, we  divide $ \Delta \int  { \cal A}^2 $
(Fig. \ref{data}c) by   $\bar{\cal M} $
(average of  the distributions in Figs. \ref{f2_s13}a and b), and multiply
by $p'$, both known functions of $s_{13}$. Then the only $s_{13}$ 
dependence of the right hand side of  Eq. 2 is in the phase motion 
$\delta(s_{13})$. From Fig. \ref{f2_s13}, the  zero at $s_{13}\sim
 0.48$ GeV$^2/c^4$ in the angular function, produces  
a singularity around this value in 
$ \Delta \int  { \cal A}^2 p'/\bar{\cal M} $. 
In Fig. \ref{data3} we show the  
$ \Delta \int  { \cal A}^2 p'/$ $\bar{\cal M} $ distribution. 
To treat the effect of the singularity, we have used a binning  such 
 that the singularity is placed in the middle of one bin. Doing  
this, we isolate the singularity in a single bin (bin 6) and discard it 
in the analysis.  We point out that the location of this 
singularity  can only affect the exact position of the minimum of $ \Delta 
\int  { \cal A}^2 p'/\bar{\cal M} $. It does not change the general 
features observed, in  Fig. \ref{data3}, that this quantity starts at zero, 
has   maximum and minimum values, 
and comes back to zero, the signature  for a  strong phase variation. 
The confidence level for a straight line fit to the data in Fig. \ref{data3} is  
4.6\%,  while the separation
between the maximum and minimum values has a significance level
of 2.6 r.m.s.

  We can see that the 6$^{\rm th}$ bin has a huge error, 
which corresponds to the bin due to the presence of the singularity.

%%%%%%%%%%%%%%%%%%%%%%%%%%%%%%%%%%%%%%%%%%%%%%%%%%%%%%%%%%%%%%%%%%%
%%%%%%%%%%%%%%%%%%%%%%%%%%%%%%%%%%%%%%%%%%%%%%%%%%%%%%%%%%%%%%%%%%%

Assuming that $\delta( s_{13})$ is an analytical
function of $ s_{13}$, Eq. 2 allows us to set the two following
conditions at the maximum and minimum values of
$ \Delta \int  { \cal A}^2  p'/ \bar{\cal M} $, respectively.

\begin{equation}
(\Delta\int{ \cal A}^2)  p'/ \bar{\cal M}_{max} \rightarrow sin(2\delta (s_{13})+\gamma) = -1
\end{equation}

and 
\begin{equation}
(\Delta\int{ \cal A}^2) p'/ \bar{\cal M}_{min} \rightarrow sin(2\delta (s_{13})+
\gamma) = 1.
\end{equation}

\noindent With these two conditions, we obtain  $\gamma$ and $\cal C$, and
 with these values we can extract directly  $\delta( s_{13})$   from 
Fig. \ref{data3} by inverting Eq. 2.

To propagate   the statistical errors  from   Fig. \ref{data}
to the values of the  $\gamma$ and $\delta( s_{13})$, we
``produce" statistically compatible ``experiments" by allowing each bin of
Fig. \ref{data}a and Fig. \ref{data}b to fluctuate randomly
following a Poisson law. We then solve the problem for each set. The 
 statistical error in each bin for  $\delta(s_{13})$ is
the r.m.s. of the  $\delta(s_{13})$ distributions  from 
the Monte Carlo experiments. For the systematic errors, we change the
$  \epsilon $   parameter ($  \epsilon = 0.22$ GeV$^2/c^4$  and $  \epsilon = 
0.30$ GeV$^2/c^4$);  we examine the possible influence of other neglected amplitudes
 contributing 
in this region of the phase space (based on the E791  amplitude measurements
for non-resonant, $f_0(1370) \pi^+$ and $\rho^0(1450) \pi^+$\cite{prl}
  contributions); and, to study the effect  of averaging 
$ ^{J=2}\bar{\cal M}_{f_2(1270)}(s_{13}) $
distributions, we  use each  distribution
 of Fig. \ref{f2_s13}a  and Fig. \ref{f2_s13}b separately. The three systematic errors, while treated 
 separately bin-by-bin, are found to be of an average size, relative to 
the statistical uncertainty, of 1, 0.6, and 0.4, respectively. They are added in 
 quadrature.
 
We measure $\gamma =  3.31 \pm  0.33 \pm 0.49 $ (with the
first error statistical and the  second  systematic)\footnote{ Here, in addition to the  
systematic error sources mentioned above, we have also incorporated one related
to the binning choice that we made to avoid the singularity in  $\bar{\cal M} $. If we change the
binning such that the singularity is placed between bins we measure 
$\gamma =  3.52 \pm  0.53$.}. The value  is
somewhat  larger than  the E791 full Dalitz-plot analysis value 
($\gamma_{Dalitz} =  2.59 \pm 0.19 $) \cite{prl}. The asymmetry of the 
distribution in Fig. \ref{f2_s12} and  the consequent use
of an effective mass-squared for the $f_2(1270)$ = 1.535 GeV$^2/c^4$ instead of
the nominal mass is  responsible for the observed shift.  To evaluate the magnitude
  of  this effect,  we generated 1000 fast MC
 samples with only two amplitudes,  $f_2(1270)$ and  
$\sigma(500)$. For both, we
used  Breit-Wigner functions with the E791 parameters, including 
 the  phase-difference of 2.59 rad. We extract   
 $\gamma$ from these 1000 samples with the method presented here.
The result has  a mean value of 3.07 $\pm$ 0.10 rad, instead of 
the input value 2.59 rad. 
We estimate an offset of  -0.48 (2.59 - 3.07) for  $\gamma$ from the 
difference  between the generated and measured values  in this Monte Carlo test.
This yields to a corrected $\gamma_{corr} =  2.83 \pm 0.38 \pm 0.49$.
The production phase-difference between the $f_2(1270)\pi^+$ and the
$\sigma(500) \pi^+$ decays of $D^+$ measured in the isobar Dalitz analysis
is in good  agreement with  $\gamma_{corr}$  from  the AD method.

With our  $\gamma$ and ${\cal C}$ values we solve Eq. 2 for $\delta(s_{13})$
for each $s_{13}$ bin. However, there are ambiguities that  arise due to  
the $sin^{-1}$ operations. Table \ref{taberr}
shows the four possible solutions for $\delta(s_{13})$\footnote{The occurrence
of two different sets of statistical  
errors in $\delta(s_{13})$ is caused by convolution of the  error in 
$\gamma$ and in $\Delta  \int {\cal A}^2$, and the fact that the 
distributions of those variables are not symmetrical. Moreover, the method 
naturally favors $\delta(s_{13})$ values coming from the maximum and 
minimum bins in Figure 7, since these are used to determine the 
$C$ and $\gamma$ that are in turn used to determine all $\delta(s_{13})$ 
values.  The 
uncertainties are therefore relatively smaller in these two bins than in the 
  other bins, a result that may be attributed in part to 
the model itself rather than the quality of the data in these bins.}.
To resolve the ambiguities we use  the assumption that  the phase 
is  zero at threshold and is an  increasing, 
monotonic, and smooth function of $s_{13}$.

\begin{table}[htb] \centering
 \begin{tabular}{|c|c|c|c|c|c|c|}     \hline
  bin  & $ \Delta \int  { \cal A}^2 \hspace{.1cm} p'/ \bar{\cal M} $&
  $\delta_0(s_{13})(^{\circ})$ & $\delta_1(s_{13})(^{\circ})$   &  $\delta_2(s_{13})(^{\circ})$ 
  &$\delta_3(s_{13})(^{\circ})$ & systematic$(^{\circ}$) 
 \\ \hline 
 {\bf 1} &$0.27\pm 0.18$&$ -104\pm 20 $&$ 76 \pm  20 $&$ {\bf  7\pm 8 }$&$  187\pm  8 $& {\bf 17}
 \\ 
 {\bf 2} & $ 1.31\pm 0.32$  &$ -139 \pm 15 $&$ {\bf 42 \pm  15} $&$ {\bf 42 \pm  12 }$&$ 222 \pm  12 $ & {\bf 20}
 \\ 
 {\bf 3} & $ 0.74\pm 0.43$ &$ -110 \pm  21$&$ {\bf 70 \pm 21 }$&$  13\pm 13 $&$ 193\pm  13 $  & {\bf 39}
 \\ 
 {\bf 4} & $ -0.46\pm 0.71$&$ -90 \pm  26$&$ {\bf 91 \pm  26} $&$  -8\pm  16 $&$ 173 \pm 16 $ & {\bf 23}
 \\ 
 {\bf 5} & $ -2.1\pm 1.3$&$ -49 \pm 16 $&$ {\bf 132\pm 16} $&$  -49\pm 21 $&$  {\bf 132\pm 21} $ & {\bf 11}
 \\ 
 {\bf 7} & $ 0.8\pm 1.8$&$ -109 \pm 27 $&$ 71 \pm  27 $&$ 12 \pm 25 $&$ {\bf  192\pm 25} $ & {\bf 27}
 \\
 {\bf 8} & $ 0.91\pm 1.01$&$ -110 \pm 23 $&$  70\pm  23 $&$  13\pm 17$&$ {\bf 193 \pm 17} $ & {\bf 11}
 \\ 
 {\bf 9} & $ 0.97\pm 0.51$&$  -112\pm 20 $&$ 68 \pm 20 $&$  15\pm 10 $&$  {\bf 195\pm  10} $ & {\bf 13}
 \\ 
 {\bf 10} & $ 0.28\pm 0.49$&$ -101 \pm 22 $&$  79\pm 22 $&$ 4 \pm 10 $&$ {\bf 184 \pm 10 }$ & {\bf 5}
  \\ \hline
\end{tabular}
\protect\caption{  Second column are the $ \Delta \int  { \cal A}^2 \hspace{.1cm} p'/ \bar{\cal M} $
values plotted in Fig.7; note the arbitrary normalization. Columns 3 to 6 are four possible solutions
for $\delta(s_{13})$: $\delta_0(s_{13})$ 
 for $ (2 \delta (s_{13})+
\gamma) $,  $\delta_1(s_{13})$ for $(2 \delta(s_{13})+
\gamma) +2\pi$, $\delta_2(s_{13})$ for $ \pi - (2 \delta(s_{13})+
\gamma) )$ and  $\delta_3(s_{13})$ for $ \pi - (2 \delta(s_{13})+
\gamma) ) +2\pi)$. The systematic errors are described in the text.}
 \label{taberr}
 \end{table}

\begin{figure}[hbt]
\centerline{\epsfig{file=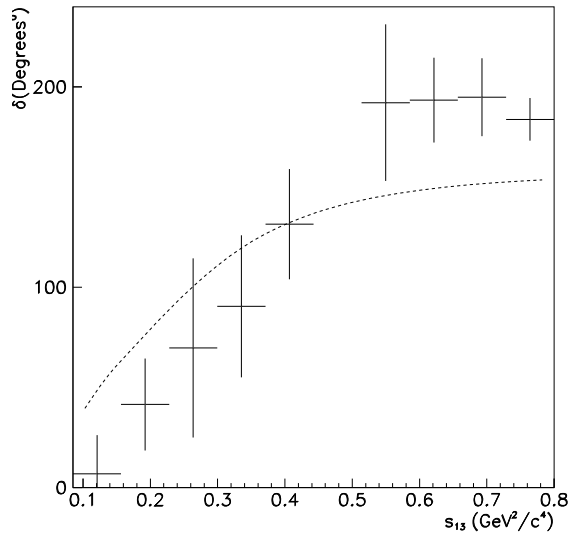,width=8 cm}}
\caption{ The phase values $\delta(s_{13})$ from our preferred solution versus the
invariant $\pi^+\pi^-$ mass squared with statistical  and systematic errors added in
quadrature. The continuous line is  the
Breit-Wigner phase motion with the E791 parameters
for the  $\sigma(500)$\cite{prl}.}
\label{phase}
\end{figure}

The solution, after the above criteria
 for $\delta(s_{13})$, including 
systematic and statistical errors, is shown in Fig. \ref{phase} and in 
bold values in Table \ref{taberr}.  We see a strong phase variation of 
about 180$^0$, starting at threshold and saturating around $s_{13} = 0.6$ 
GeV$^2/c^4$.
The limited sample size does not allow us to perform 
an accurate measurement of the mass and width parameters of a BW resonance fitted 
to this plot. We can only say that, at least for the preferred solution,  
the phase  variation  is  compatible with the complete phase 
motion through 180$^0$ expected for a resonance. This  result 
is in agreement with  the evidence 
for a broad and low mass scalar resonance suggested by the  previous E791
result using the full Dalitz-plot analysis \cite{prl}.

%%%%%%%%%%%%%%%%%%%%%%%%%%%%%%%%%%%%%%%%%%%%%%%%%%%%%%%%%%%%%%%%%%%
%%%%%%%%%%%%%%%%%%%%%%%%%%%%%%%%%%%%%%%%%%%%%%%%%%%%%%%%%%%%%%%%%%%

The Breit-Wigner phase motion for the E791 mass and width parameters,
combined with other scalar contributions obtained from that 
fit \cite{prl},
is shown as the continuous line of Fig. \ref{phase}. We can see a two 
 standard deviation difference at lowest and  highest $\pi^+\pi^-$ 
 invariant masses squared, in  opposite 
 directions, but in overall  agreement with the mass region where a scalar 
 meson $\sigma$ must have its strong phase variation. {Both results, 
 the E791 with a Breit-Wigner phase variation, and the one presented in this
 paper, show a stronger phase variation than that obtained with 
 theoretical constraints in  $\pi\pi \to \pi\pi$ elastic scattering data in the 
 scalar-isoscalar channel below 1 GeV \cite{oller,kaminski,torn3}.
 The discrepancy between these results  could be an indication that applying 
  Watson's  theorem \cite{watson} is not straight-forward when comparing the phase 
 motion of a two-body elastic interaction to the three-body decays.
 
We  have presented  an  extraction of 
the phase motion of the low mass $\pi^+\pi^-$ scalar amplitude using the 
well known $f_2(1270)$ tensor meson in the crossing  channel acting as an
interferometer.  The result  is obtained with an event counting 
procedure in a region of the phase space which is dominated by a D-wave $f_2(1270)$ 
interfering with the S-wave.   
The derivation of the phase motion relies heavily on the assumption that the 
maximum and minimum bins in Fig. \ref{data3}  correspond to the quantity
$S = sin(2\delta + \gamma) = +1$ and $-1$.  The clear presence
of a maximum and a minimum, separated from each other by 2.6 standard deviations, 
supports this assumption.  
Given this caveat, the  solution for $\delta( s_{13})$ 
has a variation of about 180$^0$, consistent with a resonant $\sigma(500)$ 
contribution. We also 
obtain good agreement between the FSI $\gamma_{corr}$ determined with the 
AD method and the $\gamma$ observed in the full 
Dalitz-plot analysis using an isobar model \cite{prl}.  These results 
support the previous  evidence for an important contribution of the 
isoscalar $\sigma(500)$ meson in the \d3pi decay \cite{prl}.
\begin {center}
{\bf Acknowledgement}
\end {center}

We would like to acknowledge the E791 Collaboration for allowing us to use their
 data for this paper and for the valuable discussions. We would like to thank  Profs.
Jeff Appel, Steve Bracker, Hans G\"unter Dosch, Brian Meadows,  Wolfgang Ochs, and 
Alberto Reis  for suggestions and important comments.

\end{document}